\documentclass[a4paper]{jpconf}
\usepackage{graphicx}
\usepackage{amsmath}
\usepackage{bm}
\usepackage[utf8]{inputenc}
\usepackage{verbatim}

\newcommand{\Rmnum}[1]{\expandafter\@slowromancap\romannumeral  #1@}

\newcommand{\ch}{\hat{c}}
\newcommand{\chd}{\hat{c}^\dagger}

\newcommand{\Hh}{\hat{H}}

\newcommand{\Gb}{ {\bf G } }
\newcommand{\Sigmab}{ {\bf \Sigma } }
\newcommand{\Gammab}{ {\bf \Gamma } }

\newcommand{\Tb}{ {\bf T } }
\newcommand{\vb}{ {\bf v } }

\newcommand{\nh}{\hat{n}}

\newcommand{\Eq}[1]{Eq. \!\!(\ref{#1})}

\newcommand{\vks}{v_{KS}}
\newcommand{\bks}{b_{KS}}

\begin{document}
\title {Effective bias and potentials in steady-state quantum transport:
A NEGF reverse-engineering study}
\author{Daniel Karlsson}
\address{Department of Physics,
Nanoscience Center P.O.Box 35 FI-40014 University of Jyväskylä, Finland 
}
\ead{daniel.l.e.karlsson@jyu.fi}

\author{Claudio Verdozzi}
\address{Department of Physics, Division of Mathematical Physics, Lund University, 22100  Lund, Sweden; and European Theoretical Spectroscopy Facility (ETSF)}
\ead{claudio.verdozzi@teorfys.lu.se}

\begin{abstract}
Using non-equilibrium Green's functions combined with many-body perturbation theory, we have calculated steady-state densities and currents through short interacting chains subject to a finite electric bias.
By using a steady-state reverse-engineering procedure, the effective potential and bias which
reproduce such densities and currents in a non-interacting system have been determined. 
The role of the effective bias  is characterised with the aid of the so-called exchange-correlation bias, recently introduced in a steady-state density-functional-theory formulation for partitioned systems.
We find that the effective bias (or, equivalently, the exchange-correlation bias) depends strongly on the interaction strength and the length of the central (chain) region. Moreover, it is rather sensitive to the level of many-body approximation used. Our study shows the importance of the effective/exchange-correlation bias out of equilibrium, thereby offering hints on how to improve the description of density-functional-theory based approaches to quantum transport.
\end{abstract}

\section{Introduction}
The method of Non-Equilibrium Green's functions (NEGF) is a general and powerful tool to describe out-of-equilibrium quantum phenomena \cite{Kadanoff1962, stefanucci2013, Balzer2013}. Nowadays, NEGF are extensively used in many areas of physics, and new fields of application emerge continually \cite{Bonitz2006}. In this work, we use NEGF to gain insight into another theoretical approach. We address conceptual aspects of Density-Functional Theory (DFT) \cite{Hohenberg1964,Kohn1965} and Time-Dependent DFT (TDDFT) \cite{Runge1984} in the context of quantum transport phenomena.

In equilibrium DFT, the particle density $n$ is the fundamental variable. There exists an invertible map between $n$ and the external potential $v$, which means that all observables are functionals of $n$. In time-dependent DFT, it is the time-dependent density $n(t)$ that plays this role. Several more extensions of DFT exists, like current DFT \cite{Vignale1987}, quantum electrodynamics DFT \cite{Tokatly2013}, and so on. All these functional theories make use of the concept of reduced quantities. Furthermore, in DFT-type theories, the interacting problem is mapped into an independent-particle one, the so-called Kohn-Sham (KS) system \cite{Kohn1965}, where the non-interacting particles experience an effective one-particle potential $v_{KS}$ (commonly referred to as the KS potential). 

A geometrical arrangement often considered in quantum-transport studies is a small (nanoscale) central region, 
where inter-particle interactions are explicitly taken into account, connected to macroscopic metallic (free-electron-like) contacts.  This setup prompts the notion of partitioned systems in quantum transport \cite{Caroli1971, Cini1980, Stefanucci2004a, Kurth2005, Verdozzi2006, Myohanen2008a} for which a TDDFT description is 
 a natural choice \cite{Stefanucci2004a, Kurth2005}.
 
\begin{figure}
 \centering
 \includegraphics[width=0.7\textwidth]{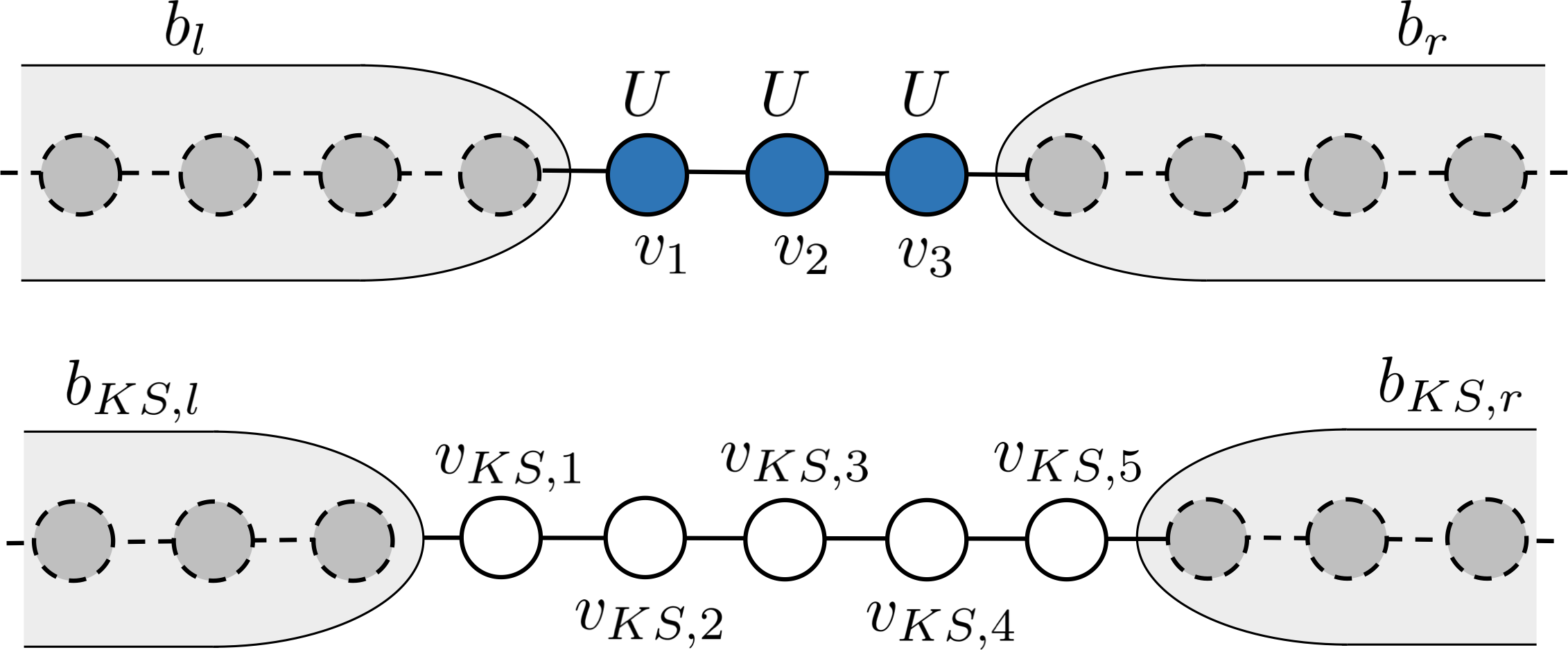}
 \caption{Schematic of the lead-chain-lead system. The upper system is the interacting system, taken for sake of illustration with three interacting sites. The lower system is the non-interacting KS system, with $v_{KS}$ and $b_{KS}$ chosen such as to obtain the same $(I,n)$ as in the interacting system. To illustrate the construction used  for steady-state TDDFT, an extended central region containing also one site from each lead is depicted. In an i-DFT description, the minimal allowed size of the central region should be the same as in the upper system.
 \label{fig:LeadChainSystem} }
\end{figure}

However, even if the interactions are confined to the central region, the exact TDDFT KS potential $v_{KS}$ is nonzero in the leads, due to the non-local nature of exchange-correlation contributions. In equilibrium $v_{KS}$ tends to zero deep in the leads, while for steady-state regimes, as reached at the end of a TDDFT time evolution (henceforth referred to as steady-state TDDFT), $v_{KS}$ tends to a constant value \cite{Stefanucci2004a}. In the latter case, the limit of $\vks$ can be interpreted as an effective bias, $\bks$.
With this perspective, one can exactly describe the system (within a given numerical accuracy) as an enlarged (with respect to the physical device) central region, including part of the leads where $v_{KS}$ is not constant in space, plus effective leads with a renormalized, but constant, bias and a current $I$ flowing through the whole system.  For a schematic, see Figure \ref{fig:LeadChainSystem}.

The above considerations can be translated into a numerical reverse-engineering scheme to find 
$\vks$ and $\bks$ in the entire system. For the aforementioned extended region, of size $L$, we gauge away the potential in, say, the right right lead, and thus we remain with $L+1$ quantities to determine ($\vks$ and $\bks$ in the left lead), using the $L$ NEGF densities in the central region and the current $I$ as constraints \cite{Preliminary}.
For this scheme, it is essential that the KS bias in the leads is approximately constant. Hence the necessity of achieving convergence by progressively increasing the size of an enlarged central region, such that all spatial variations of $v_{KS}$ are inside the region).
When converged, $\bks$ from this scheme will be $\vks$ deep  in the leads, according to the discussion above.

What just described can be rigorously justified in terms of a recently introduced DFT-like description of steady-state quantum transport \cite{Stefanucci2015}. In this formalism, the reduced quantities are the steady-state current $I$ (hence i-DFT) and the densities in the central region only. The quantities one needs to consider are the external potentials in the central region $v_c$ and the bias $b$. Close to equilibrium, i.e. for sufficiently small biases, it has been shown \cite{Stefanucci2015} that the map $(v_c,b) \to (n,I)$ is locally invertible. The map also allows for the construction of a KS system in this scheme.
In i-DFT, $v_{KS,c}$ and $\bks$ depend on the specific partitioning, even though $I$ and $n$ in the central region do not. In the central region $I$ and $n$ from steady-state TDDFT and i-DFT will agree, but the potential and bias will not. Accordingly, outside the central region the density from i-DFT will differ from the exact density. Results from
our reverse engineering procedure and from i-DFT agree when the latter is applied to an extended region large enough so 
that $v_{KS,c}$ deep in the leads and $\bks$ agree. 

In this work, we perform a reverse-engineering study of the map $(\vks,\bks) \to (n,I)$ for finite biases \cite{Preliminary}.	 Historically, determining KS potentials via numerical reverse engineering has provided valuable conceptual insight into static and time-dependent DFT approaches for continuum \cite{Almbladh1984,Lein2005}, lattice \cite{Baer2008,Li2008a,Verdozzi2008,Schmitteckert2013,Schmitteckert2008,Bergfield2012} and continuum-lattice systems \cite{Bostrom2015}. For the lattice case (the kind of systems studied here), 
the procedure, mostly applied to exact solutions from finite systems but also from NEGF treatments \cite{PuigvonFriesen2010}, can face so-called $v$-representability issues \cite{Baer2008,Li2008a,Verdozzi2008,Schmitteckert2013,Farzanehpour2012} (i.e., a map inversion may not be possible).

Additional hurdles may arise when performing reverse engineering at finite bias. For example, if the system can exhibit negative differential conductance
\footnote
{Perhaps this is most easily seen when we apply a bias to leads with finite bands, where the current becomes small if the band overlap is small.}
the inversion procedure may yield multiple solutions, in contrast to the case of the low-bias limit \cite{Stefanucci2015}.

The reverse-engineering study performed here provides us with two main results: a) Out of equilibrium, the effective bias is strongly reduced compared to the applied bias and b) the size of this correction is heavily dependent on the system parameters. 
Additionally, our study provides some insight into aspects related to i-DFT.

We begin by introducing the model, and then we describe briefly steady-state NEGF for interacting and non-interacting systems. We then describe the reverse engineering algorithm, present our results, and summarize our conclusions.

\section{Model}
We consider a finite central region contacted to macroscopic leads. All interactions are confined to the central region. The Hamiltonian of the entire system is
\begin{align}
\Hh = \Hh _{c} + \Hh _{leads} + \Hh _{cl},\label{Ham}
\end{align} 
where the three terms correspond to the central region, to the leads, and to the couplings between the central region and the leads, respectively. 

The central region consists of a small interacting 1D chain with $L$ lattice sites, described by  
\begin{align}
 \Hh _{c} = \sum_{ij \sigma }T_{ij} \chd_{i\sigma} \ch_{j\sigma}  + \sum _{i \sigma} v_i \nh_{i\sigma} + U\sum _i  \hat{n}_{i\uparrow} \hat{n}_{i\downarrow},
\label{central}
\end{align}
where $\ch_{i\sigma}$  ($\chd_{i\sigma}$) is the annihilation (creation) operator on site $i$ with spin $\sigma$, and $\nh_{i\sigma} = \chd_{i\sigma} \ch_{i\sigma}$ is the density operator. The first term describes tunneling between the sites in the chain. We choose the hopping matrix $({\bf T})_{ij} = T_{ij} = -T$ if $i$ is adjacent to $j$, and 0 otherwise. The second term is the onsite potential with elements $(\vb) _{ij} = \delta _{ij} v_i$ and the third is the interaction with strength $U$. 

The two 1D leads are semi-infinite and have the chemical potential $\mu$. Their Hamiltonian is  $\Hh_{leads}=\sum_{\alpha=r,l} \Hh_{\alpha}$, where $\alpha= r (l)$ refers to the right (left) lead, and
\begin{align}
 \Hh_{\alpha } =  \sum _{ij \in\alpha, \sigma} T^\alpha_{ij} \chd_{i\sigma} \ch_{j\sigma} + b_\alpha (t) \hat{N}_\alpha.
 \label{Hamlead}
\end{align}
$b_\alpha(t)$ is the (site-independent) bias in lead $\alpha$, and $\hat{N}_\alpha = \sum _{i \in \alpha,\sigma} \nh _{i\sigma}$ is the number operator in lead $\alpha$.  The matrix elements are $T^\alpha_{ij} = -T^\alpha$ if $i$ is adjacent to $j$, and 0 otherwise.

The lead-chain coupling is 
\begin{align}
 \Hh_{cl} = -T_{cl} \sum _\sigma \left ( \chd_{1_l \sigma} \ch_{1_c \sigma} + \chd_{1_r \sigma} \ch_{L_c \sigma}    \right )  + h.c.,
\end{align}
that is, the rightmost site in the left lead is connected via $T_{cl}$ to the leftmost site in the chain, and similarly for the right lead. 

For simplicity, we choose all tunneling parameters to be equal (transparent contacts), meaning $T=T^\alpha = T_{c l}$. We furthermore set $T=1$, which defines the unit of energy.
For our steady-state treatment, we define the applied bias $b_\alpha = b_\alpha (t\to\infty)$, and  $b = b_L - b_R$. 
Furthermore, we consider the non-magnetic case, meaning that $n_\uparrow = n_\downarrow = n$ is enforced, and we choose to work at zero temperature.

\section{NEGF in steady-state}
In this section we briefly describe NEGF in the steady state for both the interacting case and for the non-interacting KS scheme. More details will be given for the second scheme, where we also describe the reverse engineering $NEGF\rightarrow KS$ which is the novel part of this work.

\subsection{NEGF in steady-state: The interacting system}

Using a NEGF description of quantum transport for an interacting system, the retarded Green's function for our central region can be written as an $L \times L$ matrix in the site basis \cite{stefanucci2013,Myohanen2008a}, 
\begin{align}
 \Gb^R(\omega) = \frac{1}{\omega {\bf 1} - \Tb - {\bm  v} - \Sigmab^R _{MB} (\omega) - \Sigmab^R _{emb,l}(\omega) 
 - \Sigmab^R _{emb,r}(\omega)}. \label{eq:retG}
\end{align}
The so-called embedding self-energies $\Sigmab^R_{emb,\alpha}$ account in an exact way for the effects of the leads, allowing for a finite matrix description of an infinite system. All interaction effects from the central region are contained in the many-body self-energy $\Sigmab^R_{MB}$. 

In equilibrium, knowledge of $\Gb^R(\omega)$ is enough to determine properties of the system. In steady-state, we need one more quantity and here we choose the lesser Green's function $\Gb^< (\omega)$. It is directly related to the density, and is defined by the Keldysh equation 
\begin{align}
 \Gb^<(\omega) = \Gb^R(\omega) \left ( \Sigmab^< _{MB} (\omega) + \Sigmab^< _{emb,l}(\omega) + \Sigmab^< _{emb,r} (\omega),
 \right ) \Gb^A(\omega). 
 \label{eq:lesserG}
\end{align}
where $\Gb^A = (\Gb^R)^\dagger$.

As approximations to $\Sigmab_{MB}$, we will make use of two standard approximations: 2nd Born (see e.g. \cite{Karlsson2014b,Thygesen2008}) that takes into account all diagrams of the self-energy up to second order, and the particle-particle T-Matrix Approximation (TMA) \cite{Galitski1958,Friesen2009,PuigvonFriesen2011,Schlunzen2016}. In these approximations, $\Sigmab_{MB}^{R/<} = \Sigmab_{MB}^{R/<} [\Gb^R,\Gb^<]$, and thus the equations have to be solved self-consistently.  Both approximations are conserving in the Kadanoff-Baym sense \cite{Baym1961, Baym1962}, i.e. certain conservation laws are fulfilled. For us, the most important one is the continuity equation, which in the context of steady-state transport implies that the current through each lead is the same, $I_l = -I_r = I$ (we define $I_\alpha$ to be the current out from lead $\alpha$).

The embedding self-energies are given by
\begin{align}
 \Sigmab _{emb,\alpha}^< = i f(\omega - \mu - b_\alpha) \Gammab ^\alpha (\omega), \quad \Gammab^\alpha (\omega) = -2\Im m \Sigmab^R_{emb,\alpha} (\omega).
\end{align}
The matrix structure of $\Sigmab^R_{emb,\alpha}$ is given by
\begin{align*}
(\Sigmab^R_{emb,\alpha})_{ij}(\omega) = \Sigma^R _{emb} (\omega - b_\alpha) \delta _{ij}
\left ( \delta _{1i} \delta_{\alpha l} + \delta _{Li} \delta _{\alpha r} \right ). 
\end{align*}
Since the leads are semi-infinite and tight-binding, $\Sigma^R_{emb}(\omega)$ has an analytic expression \cite{stefanucci2013}. 

This defines the Green's functions of \Eq{eq:retG} and \Eq{eq:lesserG}, which are solved self-consistently. The convergence was improved by using the Pulay mixing scheme \cite{Thygesen2008,Pulay1980}. The site density is obtained directly from the lesser Green's function, while the current is obtained from the Meir-Wingreen formula \cite{Meir1992}:
\begin{align}
 n_k &=  \int _{-\infty} ^\infty \frac{d\omega}{2\pi i} (\Gb ^<)_{kk}(\omega) \label{eq:density} \\
 I &= \int _{-\infty} ^\infty \frac{d\omega}{2\pi i} 
 Tr \left [\Gammab ^L (\omega) \left ( \Gb^< (\omega) - 2\pi i f(\omega - \mu - b_L) {\bf A} (\omega)   \right )     \right ], \label{eq:meir}
\end{align}
where the spectral function $ 2\pi {\bf A} = i(\Gb^R - \Gb^A)$ and $f(\omega) = \theta(-\omega)$ is the Fermi function at zero temperature. $n$ and $I$ are then used as an input into the KS reverse engineering scheme.

\subsection{NEGF in steady-state: Kohn-Sham Green's functions}
We define the KS Green's functions as the Green's functions $\Gb_{KS}^R, \Gb_{KS}^<$ that  yield the same density and current as in the original system, but which pertain to a non-interacting system with an effective (diagonal) potential ${\bf v}_{KS}$. The defining equations are 
\begin{align}
 \Gb^R_{KS} (\omega) &= \frac{1}{\omega {\bf 1} - \Tb - {\bf v} _{KS} - \Sigmab^{R,KS} _{emb,l}(\omega) - 
 \Sigmab^{R,KS} _{emb,r}(\omega)} \\
  \Gb_{KS}^<(\omega) &= \Gb^R_{KS} (\omega) \left (\Sigmab^{<,KS}_{emb,l} (\omega) + \Sigmab^{<,KS}_{emb,r} (\omega) \right ) \Gb^A_{KS}(\omega)\\
  \Sigmab _{emb,\alpha}^{<,KS} &= i f(\omega - \mu - b_{KS,\alpha}) \Gammab_{KS}^\alpha (\omega) \\
  \Gammab^\alpha_{KS} (\omega) &= -2\Im m \Sigmab^{R,KS}_{emb,\alpha} (\omega) \\
  (\Sigmab^{R,KS}_{emb,\alpha})_{ij}(\omega) &= \Sigma^R _{emb} (\omega - b_{KS,\alpha}) \delta _{ij}
\left ( \delta _{1i} \delta_{\alpha l} + \delta _{Li} \delta _{\alpha r} \right ). 
\end{align}
The KS system are described by the same hopping matrices as the original system \cite{Verdozzi2008,Farzanehpour2012}, resulting in identical shapes for the embedding self-energies. The difference is that it is $b_{KS,\alpha}$, instead of $b$, that enters them.
\footnote{It is, however, possible to consider a KS lattice with different hopping matrices, but one can then run into severe $v-$representability issues \cite{Schmitteckert2013}, and thus we do not consider this option any further. }

Since we can add a constant potential to the KS system without changing the physical properties, we restrict, without loss of generality, $b_{KS,R} = 0$, and we define $b_{KS} = b_{KS,L}$. This fixes the gauge, which is crucial for the reverse engineering to converge. 

The densities and current from the KS system, equal to the ones from the original system, are given by the same equations as for the interacting case,
\begin{align}
 n_k &=  \int _{-\infty} ^\infty \frac{d\omega}{2\pi i} (\Gb_{KS}^<)_{kk}(\omega) \label{eq:KSdensity} \\
 I &= \int _{-\infty} ^\infty \frac{d\omega}{2\pi i} 
 Tr \left [\Gammab_{KS} ^L (\omega ) \left ( \Gb^<_{KS} (\omega) - 2\pi i f (\omega - \mu - b_{KS}) {\bf A}_{KS} (\omega)   \right )     \right ]. \label{eq:KSmeir}
\end{align}
However, since the KS system is a non-interacting one, the Meir-Wingreen formula can be written as an integral over a transmission function, 
\begin{align}
 I = \int _{\mu}^{\mu + b_{KS}} \frac{d\omega}{2\pi} \mathcal T_{KS} (\omega), \label{eq:KSmeir3}
\end{align}
where $\mathcal T_{KS}(\omega) = Tr \left [  \Gammab_{KS}^L(\omega) \Gb^R_{KS} (\omega) \Gammab^R_{KS} (\omega) \Gb^A_{KS} (\omega)    \right ]$ is the KS transmission function. At this point, we wish to emphasise that i) \Eq{eq:KSmeir3} is a non-equilibrium formula, and all Green's functions need to be evaluated at the bias $b_{KS}$ ii) \Eq{eq:KSmeir3} gives the correct current, provided that $b_{KS}$ and $v_{KS}$ exists. 

In short, the current from a NEGF treatment of the original many-body system is now recast in the form of an integral over a non-interacting transmission function $\mathcal T_{KS}$. The physical interpretation of $\mathcal T_{KS}$ is not clear though, since it is the transmission function of a fictitious image (KS) system: Physical meaning should not be attributed  to $\mathcal T_{KS}$ itself, but rather to its integral. On the other hand, for low bias $\mathcal T_{KS}$ gives the KS conductance which, as shown next, can be related to the physical conductance. 

As usual in DFT-like schemes, $v_{KS}$ can be decomposed into the external potential $v$, the Hartree potential $U n_i$, and the exchange-correlation potential $v_{xc}$. Likewise, we define the effective bias to be the sum of the external bias and the exchange-correlation bias $b_{xc}$, (this latter quantity was recently introduced in i-DFT \cite{Stefanucci2015} as one of the basic variables):

\begin{align}
 v_{KS}(i) &= v_i + U n_i + v_{xc}(i)\label{XCpots} \\\
 b_{KS} &= b + b_{xc}.
\end{align}
If $U \to 0$, then $v_{xc}, b_{xc} \to 0$. Since $\bks$ and $b_{xc}$ differ by the external bias, in the following we will use them interchangeably. The introduction of $b_{xc}$ 
allows for a rewriting of the current expression, \Eq{eq:KSmeir3}, for low bias. We assume that $\mathcal T_{KS}$ tends smoothly, for small $b_{KS}$, to the equilibrium one. This yields
\begin{align}
 \left ( \frac{d I}{d b_{KS}} \right ) _{b_{KS} \to 0}= \frac{1}{2\pi} \mathcal T _{KS} (\omega=\mu,b_{KS}\to 0) = \sigma_{KS}
\end{align}
where $\sigma_{KS}$ is usually referred to as the KS conductance. 
Using $\sigma = \frac{dI}{db}$ and the chain rule, we can relate $\sigma_{KS}$ to $\sigma$ as
\begin{align}
  \sigma = \sigma_{KS} \left ( 1+\frac{db_{xc}}{db}  \right )_{b\to0} , \label{conductance}
\end{align}
a result originally derived in i-DFT \cite{Stefanucci2015}, and discussed here in a slightly different form. 
While not operationally used in the rest of the paper, Eqs. (\ref{XCpots}-\ref{conductance}) and  the considerations above elucidate the importance of the $b_{xc}$ correction.
According to Eq. (\ref{conductance}), 
we need to apply a different bias to the KS system than the one to the physical system, i.e.
the conductance is the sum of the KS conductance and a dynamical correction \cite{Stefanucci2015, Kurth2013a}, which can be sizeable even if $b \to 0$. Correspondingly, a calculation which approximates $v_{xc}$ but neglects $b_{xc}$ (which is the same as to assume $\sigma = \sigma_{KS}$) will overestimate (assuming $b_{xc} < 0$) the calculated conductance, a scenario that occurs in NEGF+LDA calculations \cite{Kurth2013a}. This is of paramount importance
for strongly correlated systems. The latter have low conductance because of interaction effects, and  $b_{xc}$ can be on the order of $-b$, giving a vanishing physical conductance while keeping a large $\sigma_{KS}$.

 \begin{figure}
  \centering
   \includegraphics[width=0.9\textwidth]{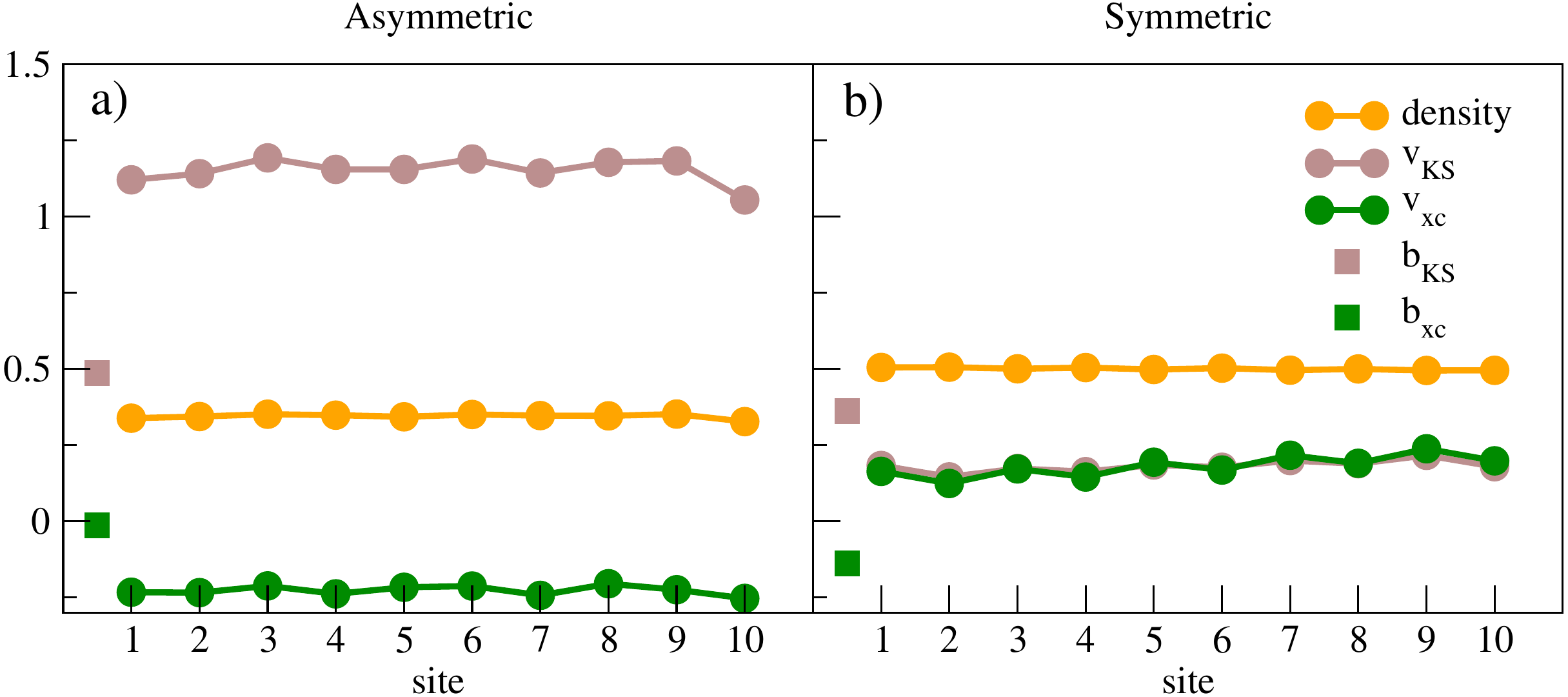}
  \caption{ Density and effective potentials and biases for a $L=10$-site chain. The applied bias is $b=0.5$, while $U=4$ and $\mu = 0$. In $a)$, $b=b_L, b_R = 0$, and $v_i = 0$. In b), $b_L=-b_R=0.25$, and $v_i = -U/2$, leading to a more homogeneous density. Since $b_{KS,R} = 0$, the zero potential is at 0.25 in b). The values of $b_{KS}$ and $b_{xc}$ are shown as squares.
  \label{fig:2} }
 \end{figure}

\subsubsection{Details regarding the reverse engineering algorithm}
The KS system should reproduce $L$ lattice densities and the current, giving $L+1$ constraints. These are intertwined, in the sense that changing $v_{KS}$ also affects the current and changing $b_{KS}$ also affects the density in the system. Thus, $\vks$ and $\bks$ have to be adjusted simultaneously in a self-consistent scheme. We found that a method that seems to be quite stable is to first fix the bias, and then change $v_{KS}$ until the densities match. At this point, we adjust $b_{KS}$, and then change $v_{KS}$ again, and so on. 

More in detail, for a fixed bias $b_{KS}$, we start from an initial $v_{KS}$. We obtain the corresponding KS Green's functions, and then calculate the density via \Eq{eq:KSdensity}. We then adjust $v_{KS}$ according to the following scheme
\begin{align}
 v_{KS} (i) \to v_{KS}(i) + \beta \frac{n^{KS}_i-n_i}{n_i},
\end{align}
where $n_i$ is the original density, and $\beta$ is a numerical parameter that can be reduced to ensure a more stable convergence. When the density is converged, we change $\bks$ via
\begin{align}
 b_{KS} \to b_{KS} - \gamma \frac{I^{KS}-I}{I},
\end{align}
where $\gamma$ plays a role similar to $\beta$. With this new $b_{KS}$, we perform another density self-consistency, and then repeat the cycle until the densities and current are converged. 

There is no guarantee that the converged solution is unique. Since the current is a non-monotonic function of the bias if we have finite leads, we expect to find at least two solutions, one for small bias, and one for high bias. In order to circumvent this problem, we start the self-consistent calculations with a low bias as initial guess.

\section{Results}
The shapes of $v_{KS}$ and $b_{KS}$ are, of course, heavily dependent on the physical parameters in the system. These quantities also depend in a non-local way on the densities and the currents. In the unbiased case, we must have $b_{KS} = 0$ since $I=0$. Out of equilibrium, based on our many-body approximations, it seems that $b_{KS} < b$ ($b_{xc} < 0$), that is, the effective bias is lessened by exchange-correlation effects. To illustrate this point, we consider an interacting chain with $L=10$ sites
subjected to two different biases, shown in Figure \ref{fig:2}. Electronic correlations are included at the 2nd Born level. In panel a), we consider an asymmetric bias with $b=b_L=0.5, b_R = 0$. The interaction keeps the density quite uniform in the chain. In b), we consider a symmetric bias $b_L = -b_R = 0.25$, and put $v_i = -U/2$ in the chain, which makes the density even more uniform. 

We see that $v_{KS}$ shows considerable more structure (especially for the asymmetric case) than the density. This is because the application  of a bias results in a potential step created at the interface between the lead(s) and the central region. In a non-interacting system, this in turn induces (Friedel-like) density oscillations, which are instead highly damped in the interacting system. For the symmetric case, Figure \ref{fig:2} b), the density is close to the lead density (half-filling). 
In this case, we find $v_{xc} \approx \vks$ since the Hartree potential is very close to $U/2$. 
Since $\bks$ is reduced compared to $b$, there is still an asymmetry in the KS system, and to reproduce the almost uniform density within the non-interacting system, the corresponding $v_{KS}$ must then oscillate in space. 
Clearly, $v_{KS}[n,I]$ depends non-locally on the density, an aspect missed by any local-density approximation.

\begin{figure}
 \centering
 \includegraphics[width=0.9\textwidth]{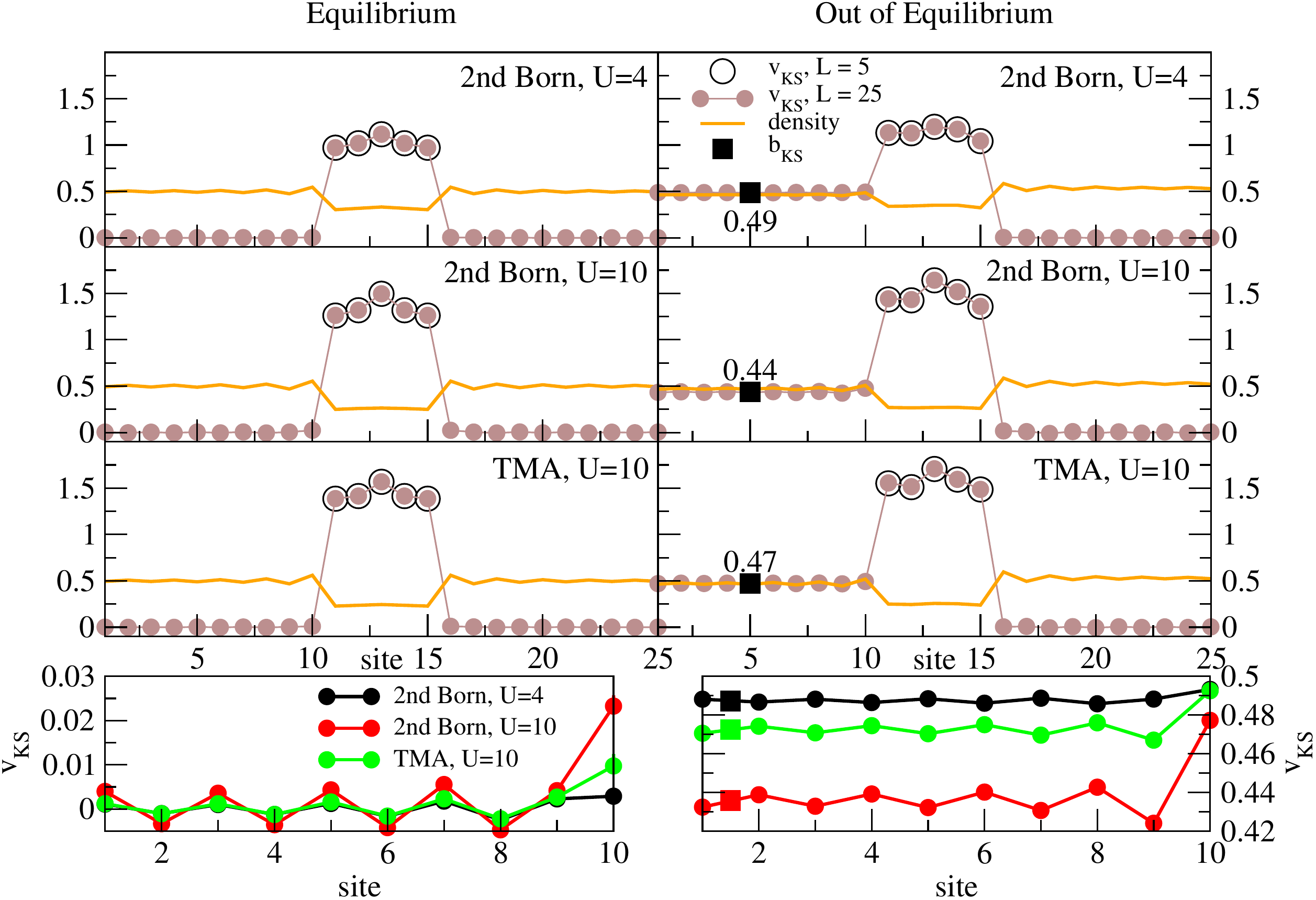}
 \caption{Reverse engineered $v_{KS}$ and $b_{KS}$ for a non-biased (left) and biased with $b_L=0.5, b_R=0$ (right) system for an interacting region of 5 sites. The leads are half-filled ($\mu = 0$). In the $L=5$ case, we do not allow for spatial variations beyond the central region, and $\vks$ is shown with open circles. In the $L=25$ case, we extend the central region to contain also 10 noninteracting sites of each lead. $\vks$ is converged already for $L=5$. The density shown is the $L=25$ density.  For $U=4$, 2nd Born and TMA are very similar, and thus we show TMA for $U=10$ only. On this scale, $\bks$ does not change when extending the central region. The values of $b_{KS}$ are shown as squares.
 \label{fig:3} }
\end{figure}

\begin{figure}
 \centering
 \includegraphics[width=1\textwidth]{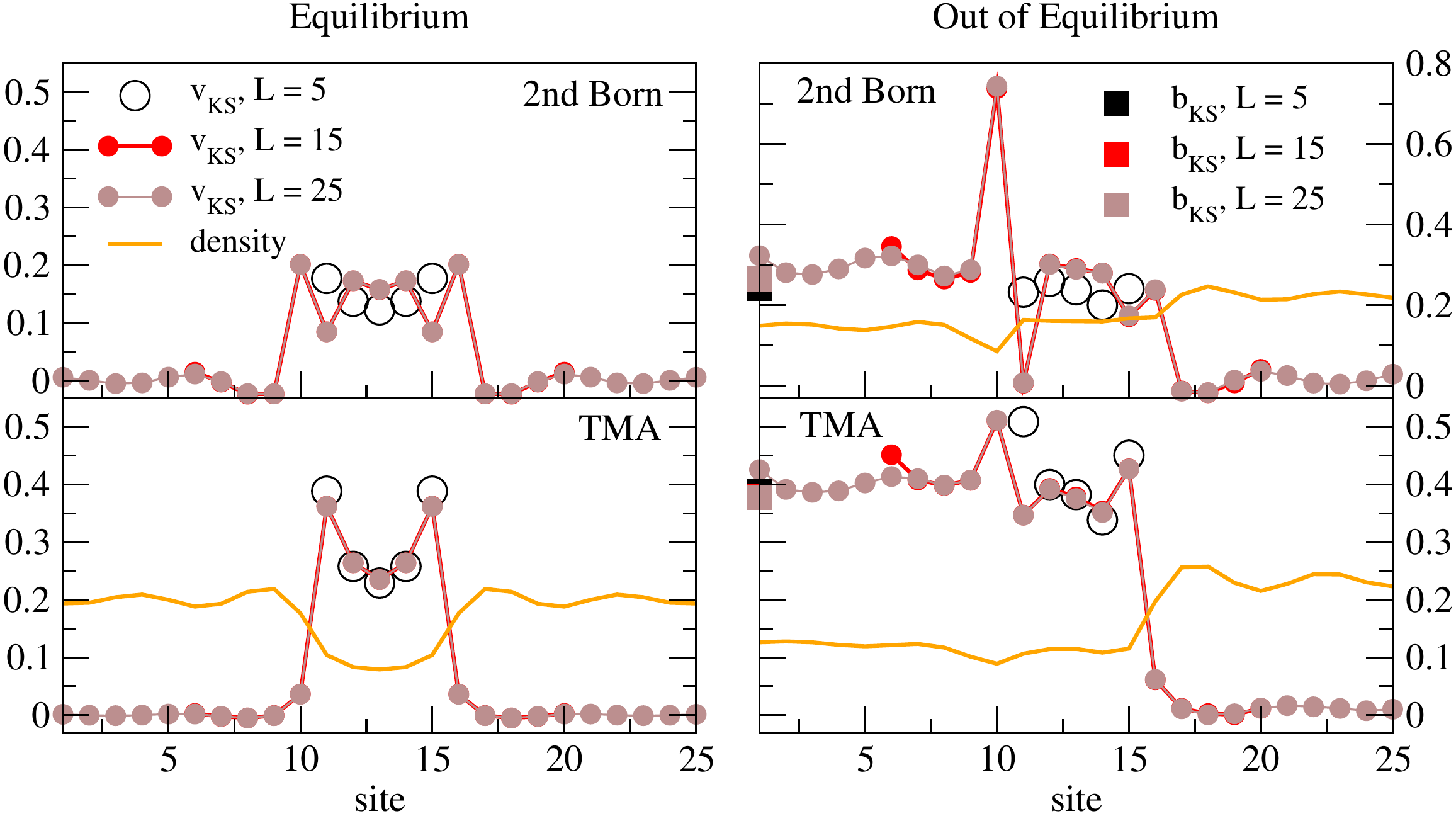}
 \caption{Reverse engineered $v_{KS}$ and $b_{KS}$ from 2nd Born and TMA for $U=10$ in the low-density regime, with average lead filling $n=0.2$ ($\mu = -2 \cos (0.2\pi)$). Allowing for a potential varying in the leads changes $\vks$ in the central region. For the unbiased case, $L=15$ is enough for $\vks$ to be converged, while $L=25$ is not enough for the biased case. Furthermore, there is a difference between $\vks$ in the lead and $\bks$. Also $\bks$ changes when increasing $L$. The density shown is for $L=25$. The values of $b_{KS}$ are shown as squares.
 \label{fig:4} }
\end{figure}

It is interesting to see how allowing for $\vks$ to vary in the leads changes $\bks$, and $\vks$ in the original central region. To illustrate the effect, we first consider a case in which $\bks,\vks$ converge very quickly (Figure \ref{fig:3}) and then one case where one needs to go to a considerably larger extended region (Figure \ref{fig:4}).  For these two situations, we take an original interacting central region of $L=5$ sites in the presence of an asymmetric bias. 

Starting with Figure \ref{fig:3}, results for a 5-site KS region (i.e. of the same size of the physical interacting region) are compared to those from an extended KS central region with additional 10 sites from each lead ($L=25$).  For the chosen set of parameters, $\vks$ in the central region is quite well converged. In the leads, $\vks$ oscillates with a decaying amplitude, tending to $\bks$ deep in the leads. This is a clear display that the effective bias is simply the effective potential deep in the leads, as shown in \cite{Stefanucci2004a}. Or, equivalently, that the exchange-correlation bias is the exchange-correlation potential deep in the leads. Finally, we note that in this parameter regime the treatment of correlations within TMA and 2nd Born yield the same qualitative behavior in the KS system.

We turn now to a set of parameters where the differences are larger (Figure \ref{fig:4}). Here, we consider
the same system as in Figure \ref{fig:3}, but instead of half-filled leads, we consider the low-density regime, $n =0.2$, where TMA is expected to perform well even for stronger interactions \cite{Galitski1958,Friesen2009,Schlunzen2016,Cini1986}.

In Figure \ref{fig:4}, we see that the choice of partitioning changes $\vks$ in the physical central region, and that there is also a difference between $\bks$ and $\vks$ at the edges of the extended region. Clearly, the largest partition considered is enough for $\vks$ to be converged in the physical central region, but not enough to reach its asymptotic value $\bks$ at the edges. 
\footnote{As function of $L$, $\vks$ at the edges of the central region and $\bks$ converge at different rates to their true KS values (the first more slowly than the second). For low lead-density, this induces a small jump between $\vks$ at the edge and $\bks$, which becomes vanishingly small only for very large $L$.  However, except for such edge sites (and immediate neighbours), $\bks$ and $\vks$ reach convergence already for moderately extended central regions.}
These considerations seem to be much more relevant for the biased case, as the unbiased scenarios are already converged for $L=15$.

\begin{figure}
 \centering
 \includegraphics[width=1\textwidth]{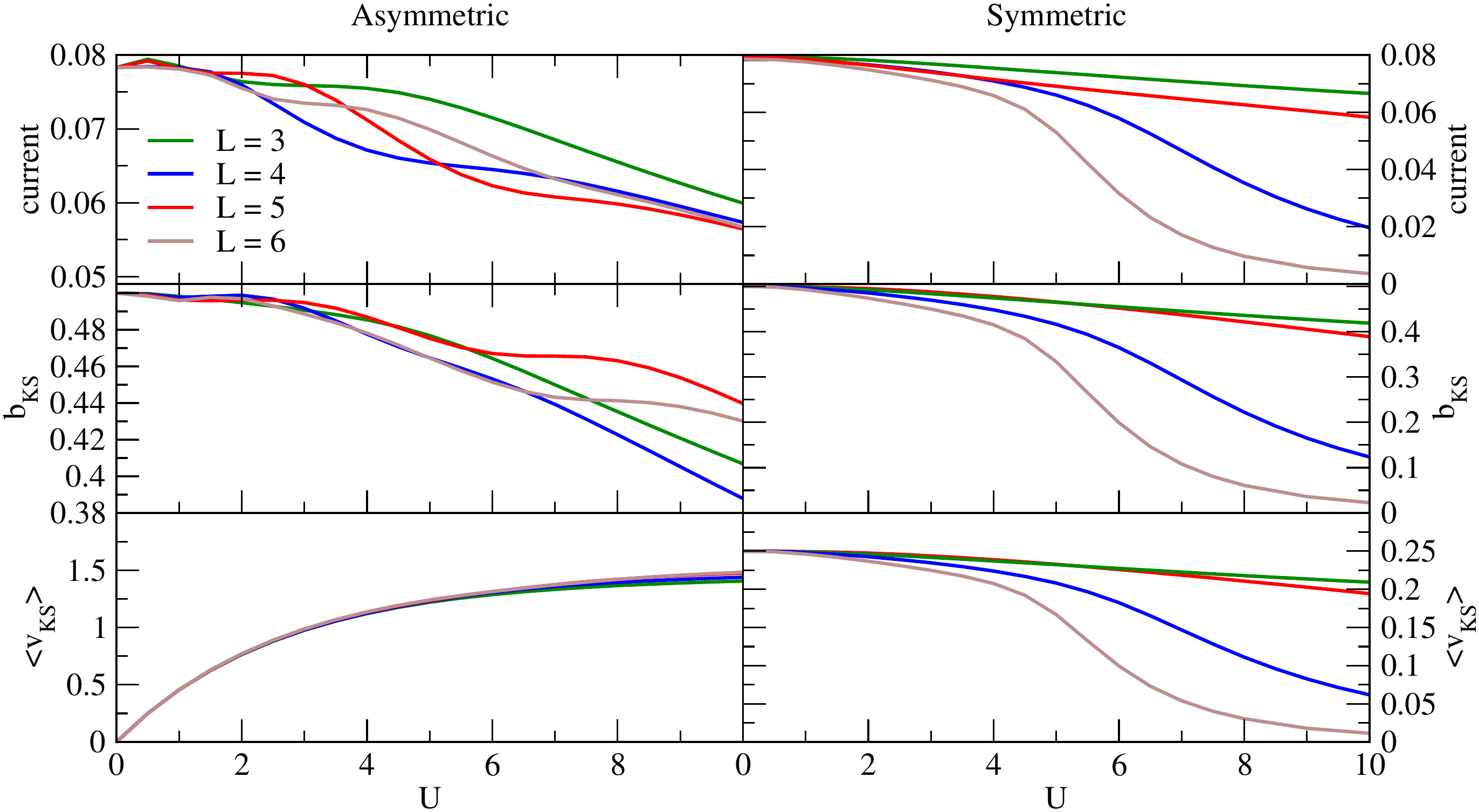}
 \caption{Currents, $b_{KS}$ and average $\vks$ for a bias $b=0.5$ for chain of various lengths, using 2nd Born. Left: $b_L = b, b_R = 0$. Right: $b_L = -b_R = b/2$ and $v_i = -U/2$.
 \label{fig:5} }
\end{figure}

We have seen that $b_{KS}$ and $\vks$ depend strongly on the system parameters. We now try to quantify this dependence by considering chains of different lengths and different interaction strengths. The results are shown in Figure \ref{fig:5}, where we also plot the  $\vks$ site-average $\langle \vks \rangle = \sum_i v_{KS,i}/L$.  We consider two different cases: asymmetric bias, and symmetric bias with $v_i = -U/2$.  

Due to interaction effects, the current generally decreases as we increase the interaction strength and/or the chain length. The KS system reproduces the same behavior by varying $\vks$ and $\bks$.  In the asymmetric case, the interplay between these two quantities induces a quite complex behavior for $\bks (U,L)$.

In the symmetric case, $\vks$ plays a slightly smaller role, and thus $\bks(U,L)$ is more similar to $I(U,L)$. Moreover, in some cases $b_{KS}$ is very small ($b_{xc} \approx -b$), showing how crucial it is to have a good description of the effective bias/exchange-correlation bias. Specific to the case of symmetric bias, we note a different pattern for odd and even numbers of sites: for even (odd) $L$ the current drops much faster (slower) as a function of $U$. These strong correlation effects are captured by the behavior of $\bks$ and $\vks$, which show a similar trend. 

Figure \ref{fig:5} shows results up to $L=6$. For longer chains with high interaction strength, numerical convergence becomes much harder to attain. However, the short chains considered here appear to be appropriate to illustrate the important trends for $\bks$ just discussed.

In this work, we have considered transparent contacts. In this case, the results indicate
that the smaller the density is, the slower $\vks$ approaches $\bks$ when moving away from the central region. On the other hand, for non-transparent contacts, more precisely for weak links and wide-band leads, the nonuniformity of $\vks$ close to the interface (on the leads side) most likely will diminish. Corroboration of this point is deferred to future work.

\section{Conclusions}
We have studied finite interacting chains strongly coupled to biased leads. We have shown that it is possible to reverse engineer effective potentials $\vks$ and biases $\bks$ that can reproduce densities and currents from many-body approximations in NEGF \cite{Preliminary}. The non-interacting Kohn-Sham system reproduces these quantities by varying $\vks$ and $\bks$ in a non-trivial way. Our results unambiguously show that effective potentials and biases depend strongly on the parameters in the system, and that it is unlikely that all correlation effects can be captured within a local-density approximation to $\vks$. We also find that  $\bks$ is reduced compared to the applied bias, in agreement with previous results in equilibrium \cite{Kurth2013a,Stefanucci2011a} and out of equilibrium regimes \cite{Stefanucci2015}. 
This further hints to the fact that any approximate KS scheme that describes transport through a strongly correlated system will need to also have a reliable approximation for $\bks$ (or $b_{xc}$), since the bias correction is sensitive to the nature of the approximation chosen. 

The NEGF currents and densities were generated by using the standard many-body approximations 2nd Born and T-matrix. While being approximate, these two different treatments of correlations produce similar trends, reinforcing our conclusions and possibly paving the way to studies to further elucidate the role of the exchange-correlation bias, by  e.g. extracting the exact $v_{KS}$ and $b_{KS}$ from steady-states obtained via time-dependent DMRG.

\ack
We acknowledge useful discussions with Miroslav Hopjan, and acknowledge the Royal Physiographic Society in Lund for support.
D.K. thanks the Academy of Finland for support.

\section*{References}
\bibliographystyle{unsrtwithoutTitle}
\bibliography{library}

\end{document}